\documentclass[%
 reprint,
showpacs,
bibnotes,
 amsmath,amssymb,
 aps,
floatfix,
altaffilletter
]{revtex4-2}

\usepackage{graphicx}
\usepackage{dcolumn}
\usepackage{bm}
\usepackage{amsmath}
\usepackage{xcolor}
\usepackage{appendix} 

\newcommand{\merav}[1]{\textcolor{black}{#1}} 
\newcommand{\nicu}[1]{\textcolor{black}{#1}} 
\newcommand{\lm}[1]{\textcolor{black}{#1}} 

\begin{document}

\title{A reservoir of timescales in random neural network}

\author{Merav Stern${^{a,b,1}}$}
\author{Nicolae Istrate${^{a,c,1}}$}
\author{Luca Mazzucato${^{a,c,d}}$}
\affiliation{${}^a$Institute of Neuroscience, ${}^c$Departments of Physics, ${}^d$Mathematics and Biology, University of Oregon, Eugene.\\ ${}^b$Faculty of Medicine, The Hebrew University of Jerusalem, Jerusalem.\\ 
${}^1$equal contribution}



\begin{abstract}
The temporal activity of many biological systems, including neural circuits, exhibits fluctuations simultaneously varying over a large range of timescales. The mechanisms leading to this temporal heterogeneity are yet unknown. Here we show that random neural networks endowed with a distribution of self-couplings, representing functional neural clusters of different sizes, generate multiple timescales of activity spanning several orders of magnitude. When driven by a time-dependent broadband input, slow and fast neural clusters preferentially entrain slow and fast spectral components of the input, respectively, suggesting a potential mechanism for spectral demixing in cortical circuits.
\\
\end{abstract}

\maketitle

\section{Introduction}
\label{sec:intro}

Experimental evidence shows that the temporal activity of many physical and biological systems exhibits fluctuations simultaneously varying over a large range of timescales. 
In condensed matter physics for example, spin glasses typically exhibit aging and relaxation effects whose timescales span several orders of magnitude \cite{bouchaud1992weak}. 
In biological systems, metabolic networks of E. coli generate fluxes with a power-law distribution \lm{of rates} \cite{almaas_barbasi2004metabolic_flux_ecoli,emmerling2002metabolic}. \lm{Gas release in yeast cultures exhibit  frequency distributions spanning} many orders of magnitude \cite{roussel2007yeast_chaos_observation}, endowing them with robust and flexible responses to the environment \cite{aon2008yeast_scalefree}. 

In the mammalian brain, a hierarchy of timescales in the activity of single neurons is observed across different cortical areas from occipital to frontal regions \cite{murray2014hierarchy,siegle2019survey,gao2020neuronal}. 
Moreover, neurons within the same local circuit exhibit a large range of timescales from milliseconds to minutes \lm{\cite{bernacchia2011reservoir,cavanagh2016autocorrelation,miri2011spatial}}. This heterogeneity of neuronal timescales was observed in awake animals during periods of ongoing activity, in the absence of external stimuli or behavioral tasks, suggesting that multiple timescales of neural activity may be an intrinsic property of recurrent cortical circuits. Recent studies highlighted the benefits of leveraging computations on multiple timescales when performing complex tasks in primates \cite{iigaya2019deviation} as well as in artificial neural networks \cite{perez2021neural}. However, the neural mechanisms underlying the emergence of multiple timescales are not yet understood. We suggest here such mechanism. 

We focus on random neuronal networks whose units are recurrently connected, with couplings that are chosen randomly. In our model, each network unit represents a functional cluster of cortical neurons with similar response properties. We interpret the unit's self-coupling \lm{as the neural cluster strength, reflecting the product of the cluster size and the average value of the recurrent synaptic coupling between its neurons}. In the case where the self-couplings are zero or weak (order $1/\sqrt{N}$), random networks are known to undergo a phase transition from silence to chaos when the variance of the random couplings exceeds a critical value \cite{sompolinsky1988chaos}. When the self-couplings are strong (order 1) and are all equal, a third phase appears featuring multiple stable fixed points accompanied by long transient activity \cite{merav2014clusters}. In all these cases, all network units exhibit the same intrinsic timescale, estimated from their autocorrelation function.
Here, we demonstrate a novel class of recurrent networks, capable of generating temporally heterogeneous activity whose multiple timescales span several orders of magnitude. We show that when the self-couplings are heterogeneous, a reservoir of multiple timescales emerges, where each unit's intrinsic timescale depends both on its own self-coupling and the network's self-coupling distribution.  \lm{In particular, we find an exponential relationship between (a power of the) unit's self-coupling and its timescales.} We analytically study the dynamics of a single unit in the limit of large self-coupling, revealing a new metastable regime described by colored noise-driven transitions between potential wells. \lm{We show that these results generalize to biologically plausible models of cortical circuits based on spiking networks with cell-type specific clustered architectures, where a reservoir of timescales emerges in presence of a heterogeneous distribution of cluster strengths}.  We then study the stimulus-response properties of our networks with heterogeneous self-couplings. In networks with zero or weak self-couplings chaotic activity is suppressed best at a single resonant frequency \cite{rajan2010stimulus}. However, when we drive our networks with a time-dependent broadband input featuring a superposition of multiple frequencies, we find that the chaotic activity is suppressed across multiple frequencies which depend on the units' respective self-couplings.
We see that units with large and small self-couplings are preferentially entrained  by  the low  and  high frequency components of the input, respectively. This spectral specificity suggests that a reservoir of timescales may be a natural mechanism for cortical circuits to flexibly demix different spectral features of complex time-varying inputs.
\begin{figure}
\begin{centering}
\includegraphics[width=0.48\textwidth]{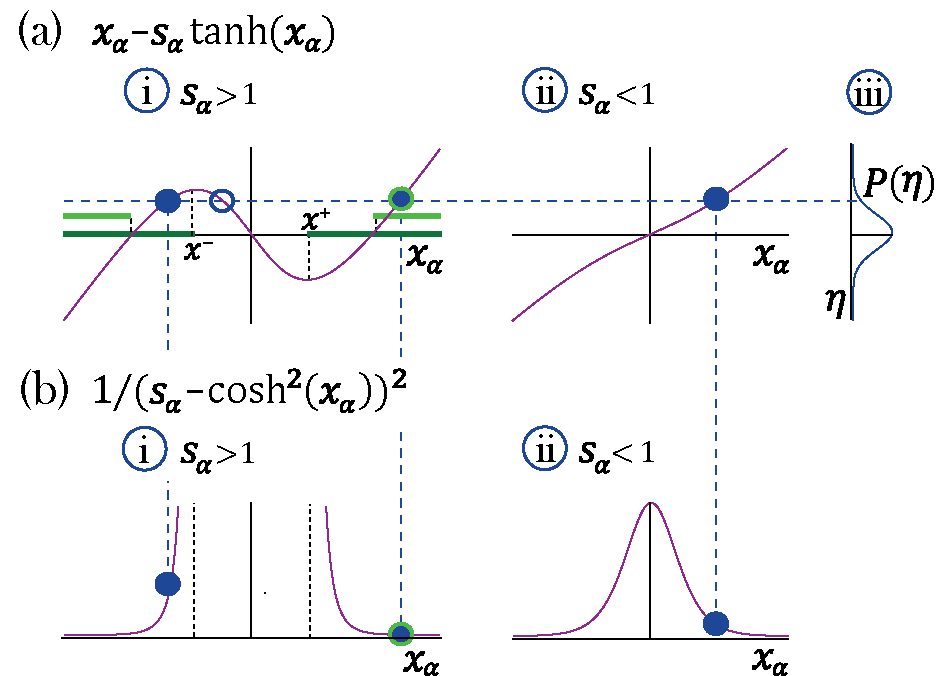}
    \caption{Transition to chaos with multiple self-couplings: Fixed point solutions and stability. a-i) The fixed point curve $x_{\alpha}-s_{\alpha}\tanh{x_{\alpha}}$, from Eq.~(\ref{eq:dmfteq1_static}), for $s_{\alpha}>1$. Stable solutions are allowed within the dark green region. b-i) The shape of a unit's contribution to stability $q^{-1}=(s_{\alpha}-\cosh{x_{\alpha}}^2)^{-2}$, from Eq.~(\ref{eq:stability}). Stable solutions of $x_{\alpha}-s_{\alpha}\tanh{x_{\alpha}} = \eta$, filled blue circles in (a-i), with different $|x|$ values contribute differently to stability. At the edge of chaos only a fixed point configuration with all units contributing most to stability (minimal $q^{-1}$) is stable, light green region in (a-i). a-ii) The curve $x_{\alpha}-s_{\alpha}\tanh{x_{\alpha}}$ for $s_{\alpha}<1$. a-iii) A possible distribution of the Gaussian mean-field $\eta$. A representative fixed point solution is illustrated by the dashed blue line: for $s_{\alpha}<1$ a single solution exists for all values of $\eta$, (filled blue circle in a-ii);For $s_{\alpha}>1$ multiple solutions exist (a-i) for some values of $\eta$; some of them lead to instability (empty blue circle in a-i). The other two solutions may lead to stability (filled blue circles in a-ii), although only one of them will remain stable at the edge of chaos (encircled with green line in a-i).} 
    \label{figone_p1}
\end{centering}
\end{figure}

\begin{figure*}
\begin{centering}
\includegraphics[width=1.0\textwidth]{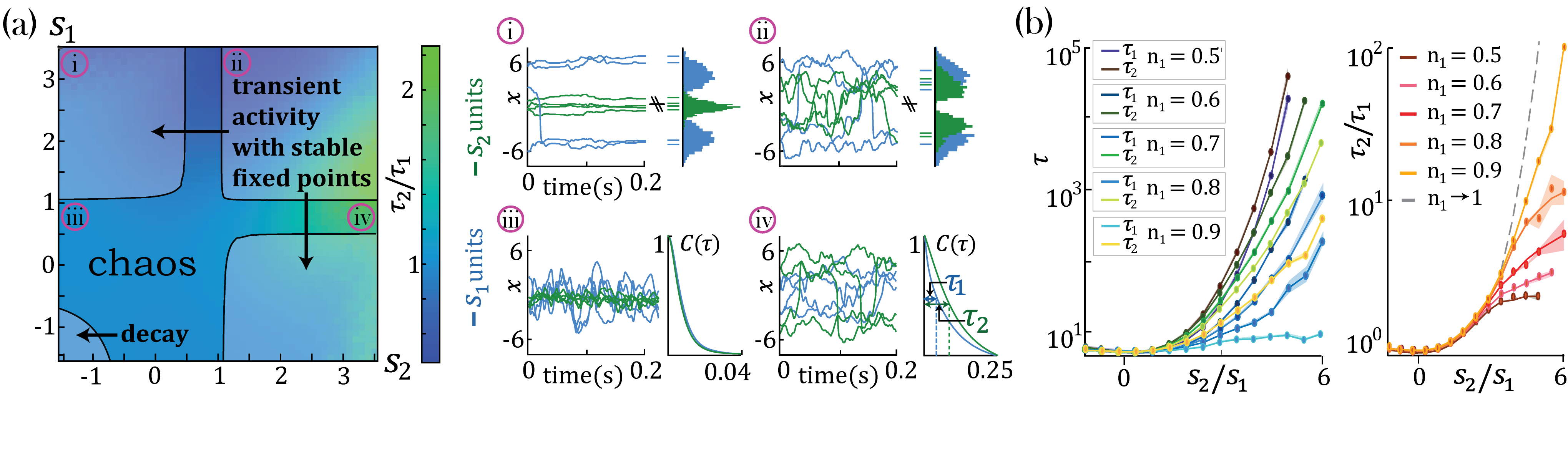}
    \caption{Dynamical and fixed point properties of networks with two self-couplings. a) Ratio of autocorrelation timescales $\tau_2/\tau_1$ of units with self-couplings $s_2$ and $s_1$, respectively ($\tau_i$ is estimated as the half width at half max of a unit's autocorrelation function, see panels iii, iv), in a network with $n_1=n_2=0.5$ and $g=2$ and varying $s_1,s_2$. A central chaotic phase separates four different stable fixed point regions with or without transient activity. Black curves represent the transition from chaotic to stable fixed point regimes. i),ii) Activity across time during the initial transient epoch (left) and distributions of unit values at their stable fixed points (right), for networks with $N=1000$ and (i) $s_1=3.2,s_2=-1.5$, (ii) $s_1=3.2,s_2=1.2$. iii),iv) Activity across time (left) and normalized autocorrelation functions $C(\tau)/C(0)$, (right) of units with (iii) $s_1=0.8,s_2=-1.5$, (iv) $s_1=0.8,s_2=3.2$. b) \lm{Timescales $\tau_2,\tau_1$ (left) and their ratio $\tau_2/\tau_1$ (right) for fixed $s_1=1$ and varying $s_2$, as a function of the relative size of the two populations $n_1=N_1/N, n_2=N_2/N$ (at $g=2$, $N=2000$; average over $20$ network realizations. }\merav{The case of relative size $n_1 \rightarrow 1$ is described in Section \ref{sec:probe} with $n_2$ relative amount of probe units. Its timescales ratio is calculated using Eq.~(\ref{large_s_timescale}) (grey dashed line).}}
    \label{figone_p2}
\end{centering}
\end{figure*}


\section{Random networks with heterogeneous self-couplings}
We consider a recurrent network of $N$ rate units obeying the dynamical equations
\begin{equation}
\frac{dx_i}{dt} = -x_i + s_i \phi(x_i) + g \sum_{j=1}^{N} J_{ij} \phi(x_j)
\label{dyn}
\end{equation}
where the random couplings $J_{ij}$ from unit $j$ to unit $i$ are drawn independently from a Gaussian distribution with mean $0$ and variance $1/N$; $g$ represents the network gain and we chose a transfer function $\phi(x) \equiv \tanh(x)$. We measure time in units of 1 ms. The self-couplings $s_i$ are drawn from a distribution $P(s)$. The special case of equal self-couplings ($s_i=s$) was studied in \cite{merav2014clusters} and a summary of the results are brought in the Appendix A for convenience. Here
we study the network properties in relation to both discrete and continuous distributions $P(s)$. 

Using standard methods of statistical field theory \cite{buice2013beyond,helias2020statistical}, in the limit of large $N$ we can average over realizations of the disordered couplings $J_{ij}$ to derive a set of self-consistent dynamic mean field equations for each population of units $x_\alpha$ with self-coupling strengths $s_{\alpha} \in S$
\begin{equation}
\label{dmfteq1}
    \frac{dx_{\alpha}}{dt}=-x_{\alpha}+s_{\alpha}\tanh(x_{\alpha})+\eta(t)\ .
\end{equation}
In our notation, $S$ denotes the set of different values of self-couplings $s_\alpha$, indexed by $\alpha\in A$, and we denote by $N_\alpha$ the number of units with the same self-coupling $s_\alpha$, and accordingly by $n_\alpha=N_\alpha/N$ their fraction. The mean field $\eta(t)$ is the same for all units and has zero mean $\langle \eta(t)\rangle=0$ and autocorrelation
\begin{eqnarray}
    \langle \eta(t)&\eta&(t+\tau)\rangle=g^2C(\tau)\nonumber\\
    C(\tau)&=&\sum_{\alpha\in A } n_{\alpha} \langle \phi[x_{\alpha}(t)]\phi[x_{\alpha}(t+\tau)]\rangle \ ,     
    \label{dmfteq2}
\end{eqnarray}
where $\langle\cdot\rangle$ denotes an average over the mean field.  
\section{Stable fixed-points and transition to chaos}
Networks with heterogeneous self-couplings exhibit a complex landscape of fixed points
$x^*_\alpha$, obtained as the self-consistent solutions to the static version of Eq.~(\ref{dmfteq1}) and Eq.~(\ref{dmfteq2}), namely
\begin{eqnarray}\label{eq:dmfteq1_static}
    x_{\alpha}-s_{\alpha}\tanh(x_{\alpha})=\eta \ ,
\end{eqnarray}
where the mean field $\eta$ has zero mean \lm{and its variance is given by}
\begin{eqnarray} \label{eq:dmfteq2_static}
    \langle \eta^2 \rangle &=& g^2 C \nonumber \\
    C &=& \sum_{\alpha\in A} n_{\alpha}  \langle \phi[x_{\alpha}]^2 \rangle \ .
\end{eqnarray}
The solution for each unit depends on its respective $s_\alpha$ (Fig. \ref{figone_p1}). If $s_\alpha<1$ a single interval around zero is available. For $s_\alpha>1$, for a range of values of $\eta$, $x^*_\alpha$ can take values in one of three possible intervals. However, the available solutions in the latter case are further restricted by stability conditions.

We can derive the stability condition by expanding the dynamical equations (\ref{dyn}) around the fixed point and requiring that all eigenvalues of the corresponding stability matrix are negative. \lm{To determine the onset of instability we look for conditions such that at least one eigenvalue develops a positive real part.} An eigenvalue of the stability matrix exists at a point $z$ in the complex \lm{plane} if \cite{merav2014clusters,ahmadian2015eigenvalues}
\begin{equation}
    g^2 \sum_{\alpha\in A} n_{\alpha} \left\langle \frac{\left[1-\tanh^2(x_\alpha)\right]^2}{\left[z + 1- s_{\alpha} \left(1-\tanh^2(x_\alpha)\right)\right]^2}\right\rangle > 1 .
    \label{eq:zstability}
\end{equation}
Since the denominator of the expression above is $z$ plus the slope of the curve in Fig. \ref{figone_p1}a-i, a solution whose value $x^*_\alpha$ gives a negative slope (available when $s_\alpha>1$) leads to a vanishing value of the denominator at some positive $z$ and hence to a positive eigenvalue and instability. Hence, the $n_\alpha$ fraction of units with $s_\alpha>1$ at a stable fixed point are restricted to have support on two disjoint intervals $[x^*_{\alpha}(s_{\alpha})<x_{\alpha}^-(s_{\alpha})]\cup[x^*_{\alpha}(s_{\alpha})>x_{\alpha}^+(s_{\alpha})]$. We refer to this regime as {multi-modal}, a direct generalization of the stable fixed points regime found in \cite{merav2014clusters} for a single self-coupling $s>1$, characterized by transient dynamics leading to an exponentially large number of stable fixed points. For the $n_\alpha$ portion of units with $s_\alpha<1$, the stable fixed point is supported by a single interval around zero.     

A fixed point solution becomes unstable as soon as an eigenvalue occurs at $z=0$, obtaining from Eq. (\ref{eq:zstability}) the stability condition
\begin{equation}
    g^2 \sum_{\alpha\in A} n_{\alpha} \langle q^{-1}_{\alpha}\rangle \leq 1
    \label{eq:stability} \ , 
\end{equation}
where $q_\alpha=\left[s_{\alpha} -\cosh^2(x_\alpha)\right]^2$. For $s_\alpha>1$ the two possible consistent solutions to (\ref{eq:dmfteq1_static}) that may result in a stable fixed point  (from the two disjoint intervals in Fig. \ref{figone_p1}a-i), contribute differently to $q_\alpha$. Larger $|x_\alpha^*|$ decreases $q_\alpha^{-1}$ (Fig. \ref{figone_p1}b-i), thus improving stability. Choices for distributions of $x_\alpha^*$ along the two intervals become more restricted as $g$ increases or $s_{\alpha}$ decreases, since both render higher values for the stability condition, Eq.~\ref{eq:stability}, forcing more solutions of $x_i$ to decrease $q_\alpha^{-1}$. This restricts a larger fraction of $x_\alpha^*$ at the fixed points to the one solution with higher absolute value. At the transition to chaos, a single last and most stable solution exists with all $x_i$ values chosen with their higher absolute value $x_\alpha^*$ (Fig. \ref{figone_p1}a-i, light green segments). For those with $s_\alpha<1$ only one solution is available, obtained by the distribution of $\eta$ through consistency (\ref{eq:dmfteq1_static}) at the fixed point. In this configuration, the most stable solution is exactly transitioning from stability to instability where (\ref{eq:stability}) reaches unity. Hence the transition from stable fixed points to chaos occurs for a choice of $g$ and $P(s)$ such that solving consistently (\ref{eq:dmfteq1_static}) and (\ref{eq:dmfteq2_static}) leads to saturate the stability condition (\ref{eq:stability}) at one.

We illustrate the discussion above in the case of a network with two sub-populations with $n_1$ and $n_2=1-n_1$ portions of the units with self-couplings $s_{1}$ and $s_2$, respectively. In the $(s_1,s_2)$ plane, this model gives rise to a phase portrait with a single chaotic region separating four disconnected stable fixed-point regions (Fig. \ref{figone_p2}a). A unit's activity is determined by its own self-coupling, the network's distribution of self-couplings and $g$. \lm{We will first discuss the stable fixed points, which present qualitatively different structures depending on the values of the self-couplings. When both self-couplings $s_1,s_2<1$, the only possibility for a stable fixed point is the trivial solution, with all $x_i=0$ (Fig. \ref{figone_p2}a), where the network activity quickly decays to zero. When at least one self-coupling in greater than one, there are three stable fixed point regions (Fig. \ref{figone_p2}a-i, a-ii, a-iii); in these three regions, the network activity starting from random initial conditions unfolds via a long-lived transient periods, then it eventually settles into a stable fixed point. This transient activity with late fixed points is a generalization of the network phase found in \cite{merav2014clusters}. When both self-couplings are greater than one ($s_1,s_2>1$) the fixed point distribution in each sub-population is bi-modal (Fig. \ref{figone_p2}a-ii,iii).} When $s_1>1$ and $s_2<1$, the solutions for the respective sub-populations are localized around bi-modal fixed points and around zero, respectively (Fig. \ref{figone_p2}a-i).  In the case of a Gaussian distribution of self-couplings in the stable fixed point regime, a complex landscape of stable fixed points emerges. The unit values at the stable fixed points  continuously interpolates between the zero (for units with $s_i<1$) and the multiple values bi-modal cases (for units with $s_i>1$) within the same network  (Fig. \ref{figtwo}a).

\section{A reservoir of timescales}
In the chaotic phase we can estimate the intrinsic timescale $\tau_i$ of a unit $x_i$ from its autocorrelation function $C(\tau) = \langle \phi[x_i(t)]\phi[x_i(t+\tau)]\rangle_t$ as the half width at its autocorrelation half maximum (Fig. \ref{figone_p2}a-vi, $\tau_1$ and $\tau_2$). The chaotic phase in the network, Eq.~(\ref{dyn}), is characterized by a large range of timescales that can be simultaneously realized across the units with different self-couplings. \lm{In a network with two self-couplings $s_1$ and $s_2$ in the chaotic regime, we found that the ratio of the timescales $\tau_2 / \tau_1$ increases as we increase the self-couplings ratio $s_2 / s_1$ (Fig. \ref{figone_p2}b). The separation of timescales depends on the relative fraction $n_2/n_1$ of the slow and fast populations.} \merav{ When this fraction approaches zero, (with $n_1 \rightarrow \infty$), the log of the timescale ratio exhibits a supralinear dependence on the self-couplings ratio, as described analytically in Section \ref{sec:probe}, leading to a vast separation of timescales. Other self-couplings ratios $s_2 / s_1$ approach the timescale supralinear separation as the self-couplings ratio increases, up to a saturation and decay back to one (not shown).}

\begin{figure}
\begin{centering}
\includegraphics[width=0.48\textwidth]{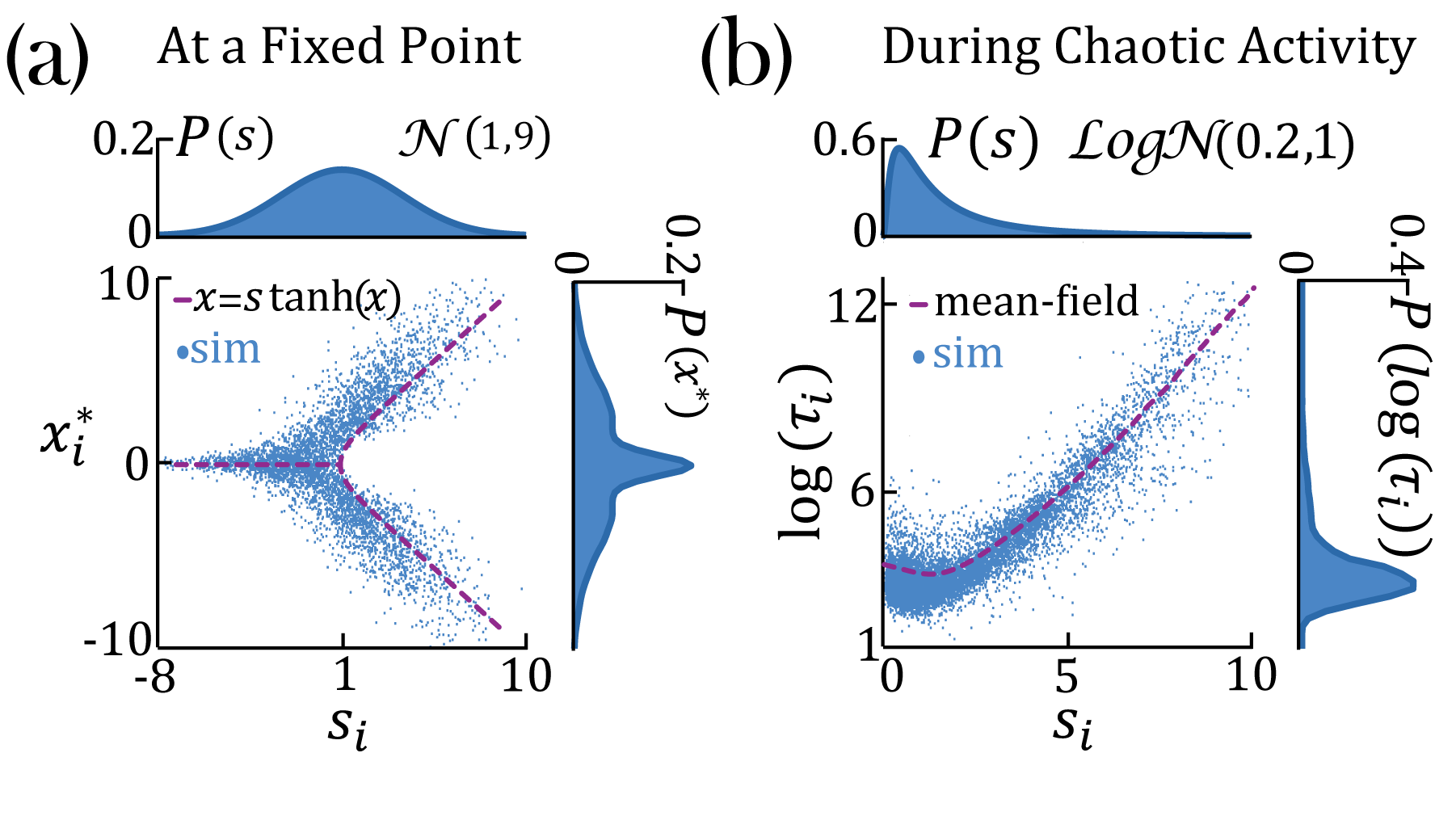}
    \caption{ Continuous distributions of self-couplings. a) In a network with a Gaussian distribution of self-couplings (mean $\mu = 1$ and variance $\sigma^2= 9$), and $g=2$, the stable fixed point regime exhibits a distribution of fixed point values interpolating between around the zero fixed point (for units with $s_i\leq1$) and the multi-modal case (for units with $s_i>1$). The purple curve represents solutions to $x = s\tanh(x)$. b) In a network with a lognormal distribution of self-couplings (parameters $\mu = 0.2$ and $\sigma^2= 1$), and $g=2$, autocorrelation timescales $\tau_i$ in the chaotic phase span several orders of magnitude as functions of the units' self-couplings $s_i$ (purple curve shows the dynamic mean-field predictions for $\tau_i$).}
    \label{figtwo}
\end{centering}
\end{figure}

In a case of a lognormal distribution of self-couplings, in the chaotic regime the network generates a reservoir of multiple timescale $\tau_i$ of chaotic activity across network units, spanning across several orders of magnitude (Fig. \ref{figtwo}b). For long tailed distributions such as the lognormal, mean field theory can generate predictions for rare units with large self-couplings from the tail end of the distribution by solving (\ref{dmfteq1}) and the continuous version of (\ref{dmfteq2}), see Appendix B, highlighting the exponential relation between a unit’s self-coupling and its autocorrelation decay time.

\begin{figure*}
\includegraphics[width=1\textwidth]{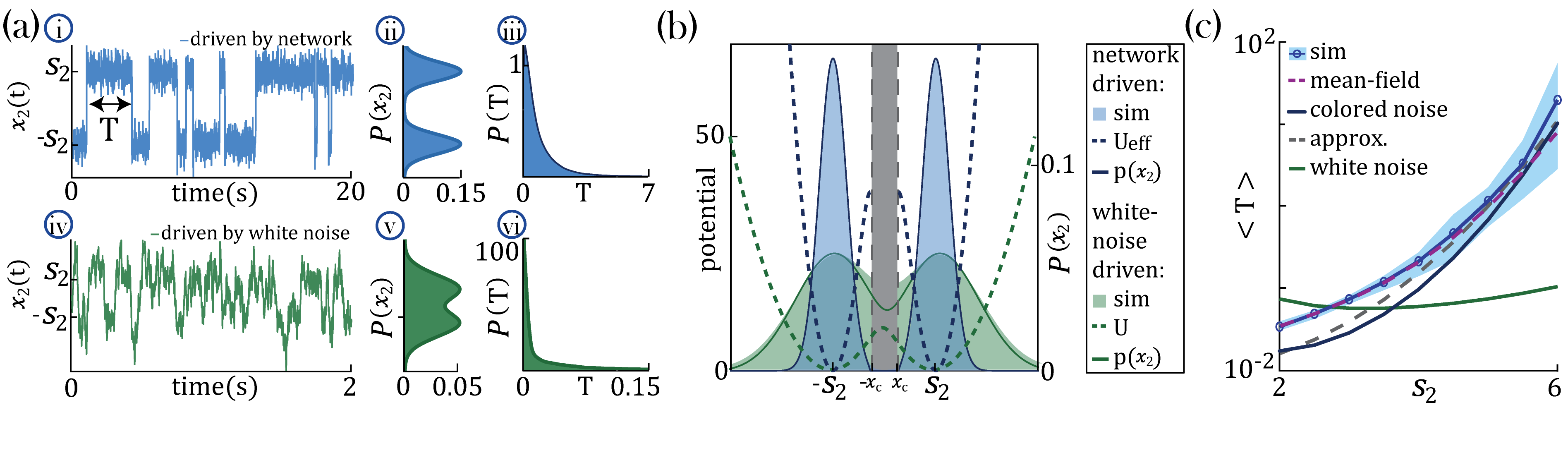}
    \caption{Separation of timescales and metastable regime. (a) Examples of bistable activity. i,iv - time courses; ii,v - histograms of unit's value across time; iii,vi - histograms of dwell times. (a-i,ii,iii) An example of a probe unit $x_2$ with $s_2=5$, embedded in a neural network with $N=1000$ units, $N_1=N-1$ units with $s_1=1$ and $g=1.5$. (a-iv,v,vi) An example of a probe unit driven by white noise. (b) The unified colored noise approximation stationary probability distribution $p(x_2)$ (dark blue curve, Eq. (\ref{pucna1}), its support excludes the shaded gray area) from the effective potential $U_{eff}$ (dashed blue curve) captures well the activity histogram (same as (a-ii)); whereas the white noise distribution $p(x_2)$ (dark green curve, obtained from the naive potential $U$, dashed green curve) captures the probe unit's activity (same as (a-v)) when driven by white noise, and deviates significantly from the activity distribution when the probe is embedded in our network. (c) Average dwell times,$\langle T\rangle$, in the bistable states. Simulation results, mean and 95\% CI (blue curve and light blue background, respectively; An example for the full distribution of $T$ is given in (a-iii)). Mean-field prediction (purple curve). The mean first passage time from the unified colored noise approximation (Eq.~(\ref{mfpt}), black curve) captures well the simulated dwell times. An approximation for the unified colored noise (Eq.~(\ref{large_s_timescale}), gray dashed line) achieves good results as well. the white noise average dwell times are significantly different (green curve).}
    \label{metastable1}
\end{figure*}

\section{Separation of timescales in the bistable chaotic regime}
\label{sec:probe}

To gain an analytical understanding of the parametric separation of timescales in networks with heterogeneous self-couplings, we consider the special case of a network with two self-couplings where a large sub-population ($N_1=N-1$) with $s_1=1$ comprises all but one slow probe unit, $x_2$, with large self-coupling $s_2 \gg s_1$. The probe unit obeys the dynamical equation
$dx_2/dt=f(x_2)+\eta(t)$, with $f(x)=-x+s_2\phi(x)$. In the large $N$ limit, we can neglect the backreaction of the probe unit on the mean field and approximate the latter as an external Gaussian colored noise $\eta(t)$ with autocorrelation $g^2 C(\tau)=g^2 \langle \phi[x_1(t)]\phi[x_1(t+\tau)]\rangle$, independent of $x_2$. The noise $\eta(t)$ can be parameterized  by its strength, defined as $D=\int_0^\infty d\tau\,C(\tau)$ and its timescale (color) $\tau_1$. 
For large $s_2$, the  dynamics of the probe unit $x_2$ can be captured by a bi-stable chaotic phase whereby its activity is localized around the critical points $x_2=x^{\pm}\simeq\pm s_2$ (Fig. \ref{metastable1}a-i) and switches between them at random times. In the regime of strong colored noise (as we have here, with $\tau_1\simeq 7.9 \gg1$), the stationary probability distribution $p(x)$  (for $x \equiv x_2$, Fig. \ref{metastable1}a-ii,b) satisfies the unified colored noise approximation to the Fokker Planck equation \cite{hanggi1995colored,jung1987dynamical}:
\begin{equation}\label{pucna1}
    p(x)=Z^{-1}|h(x)|\exp\left[-U_{eff}(x)/ D\right] \ ,
\end{equation}
where $Z$ is a normalization constant, $h(x)\equiv 1-\tau_1 f'(x)$, and the effective potential $U_{eff}(x)=-\int^x f(y)h(y)dy$ is given by
\begin{eqnarray}
U_{eff}=\frac{x^2}{2}-s_2\log\cosh(x)+\frac{\tau_1}{2} f(x)^2 -U_{min} \ .
\end{eqnarray}
The distribution $p(x)$ has support in the region $h(x)>0$ comprising two disjoint intervals $|x|>x_c$ where $\tanh(x_c)^2=1-\frac{1+\tau_1}{\tau_1 s_2}$ (Fig. \ref{metastable1}b). $p(x)$ is concentrated around the two minima $x^\pm\simeq\pm s_2$ of $U_{eff}$. The main effect of the strong color $\tau_1\gg1$ is to sharply decrease the variance of the distribution around the minima $x^\pm$. This is evident from comparing the colored noise with a white noise, when the latter is driving the same bi-stable probe $dx_2/dt = -x_2 + s_2 \phi(x_2) + \xi(t) $, where $\xi(t)$ is a white noise with an equivalent strength to the colored noise, Fig. \ref{metastable1}a-iv,v,vi. The naive potential for the white noise case $U=x^2/2-s_2\log\cosh(x)$ is obtained from (\ref{pucna1}) by sending $\tau_1\to0$ in the prefactor $h$ and in potential $U_{eff}$. It results in wider activity distribution compared to our network generated colored noise, in agreement with the simulations, Fig. \ref{metastable1}a,b.

In our network generated colored noise the probe unit's temporal dynamics is captured by the mean first passage time $\langle T\rangle$ for the escape out of the potential well:
\begin{eqnarray}\label{mfpt}
    \langle T\rangle &=&\int_{-s_2}^{-x_c}\frac{dx}{D} \frac{h(x)^2}{p(x)}\int_{-\infty}^x p(y)dy \nonumber\\
    &\simeq&2\pi\sqrt{U''_{eff}(x_-)\rho''(x_f)}\exp\left(\frac{\Delta}{D}\right) \ ,
\end{eqnarray}
where $\Delta =\rho(x_f)-U_{eff}(x_-)$ and $\rho=U_{eff}+D\log h$. We evaluated the integrals by steepest descent around $x^-$ and $-x_f$,  where $\tanh(x_f)^2\simeq 1-1/2s_2$. 
The agreement between (\ref{mfpt}) and simulation results improves with increasing $s_2$, as expected on theoretical ground \cite{hanggi1995colored,jung1987dynamical}(Fig. \ref{metastable1}c). The asymptotic scaling for large $s_2$ is
\begin{eqnarray}
    \log(\langle T\rangle)\sim \frac{\tau_1+1}{2D}\left[s_2^2-s_2\log(s_2)\right] \ .
    \label{large_s_timescale}
\end{eqnarray}

In this slow probe regime, we thus achieved a parametric separation of timescales between the population $x_1$, with its intrinsic timescale $\tau_1$, and the probe unit $x_2$ whose activity fluctuations exhibit two separate timescales: the slow timescale $T$ of the bistable switching and the fast timescale $\tau_1$ of the fluctuations around the metastable states (obtained by expanding the dynamical equation around the meta-stable values $x^{\pm}=\pm s_2$). One can generalize this metastable regime to a network with $N-p$ units which belong to a group with $s_1=1$ and $p\ll N$ slow probe units $x_\alpha$, for $\alpha=2,\ldots,p+1$, with large self-couplings $s_\alpha$. The slow dynamics of each probe unit $x_\alpha$ is captured by its own bistable switching time $T_\alpha$ in (\ref{mfpt}) and all slow units are driven by a shared external colored noise $\eta(t)$ with timescale $\tau_1$. In summary, in our model multiple timescales can be robustly generated with specific values, varying over several orders of magnitude.  

\lm{Is the relationship between the unit's self-coupling and its timescale relying on single-unit properties, or does it rely on network effects? To answer this question, we compare the dynamics of a unit when driven by a white noise input vs. the self-consistent input generated by the rest of the recurrent network (i.e., the mean field). If the neural mechanism underlying the timescale separation was a property of the single-cell itself, we would observe the same effect regardless of the details of the input noise. We found that the increase in the unit's timescale as a function of $s_2$ is absent when driving the unit with white noise, and it only emerges when the unit is driven by the self-consistent mean field. We thus concluded that this neural mechanism is not an intrinsic property of a single unit but requires the unit to be part of a recurrently connected network.}

\begin{figure}
\includegraphics[width=0.5\textwidth]{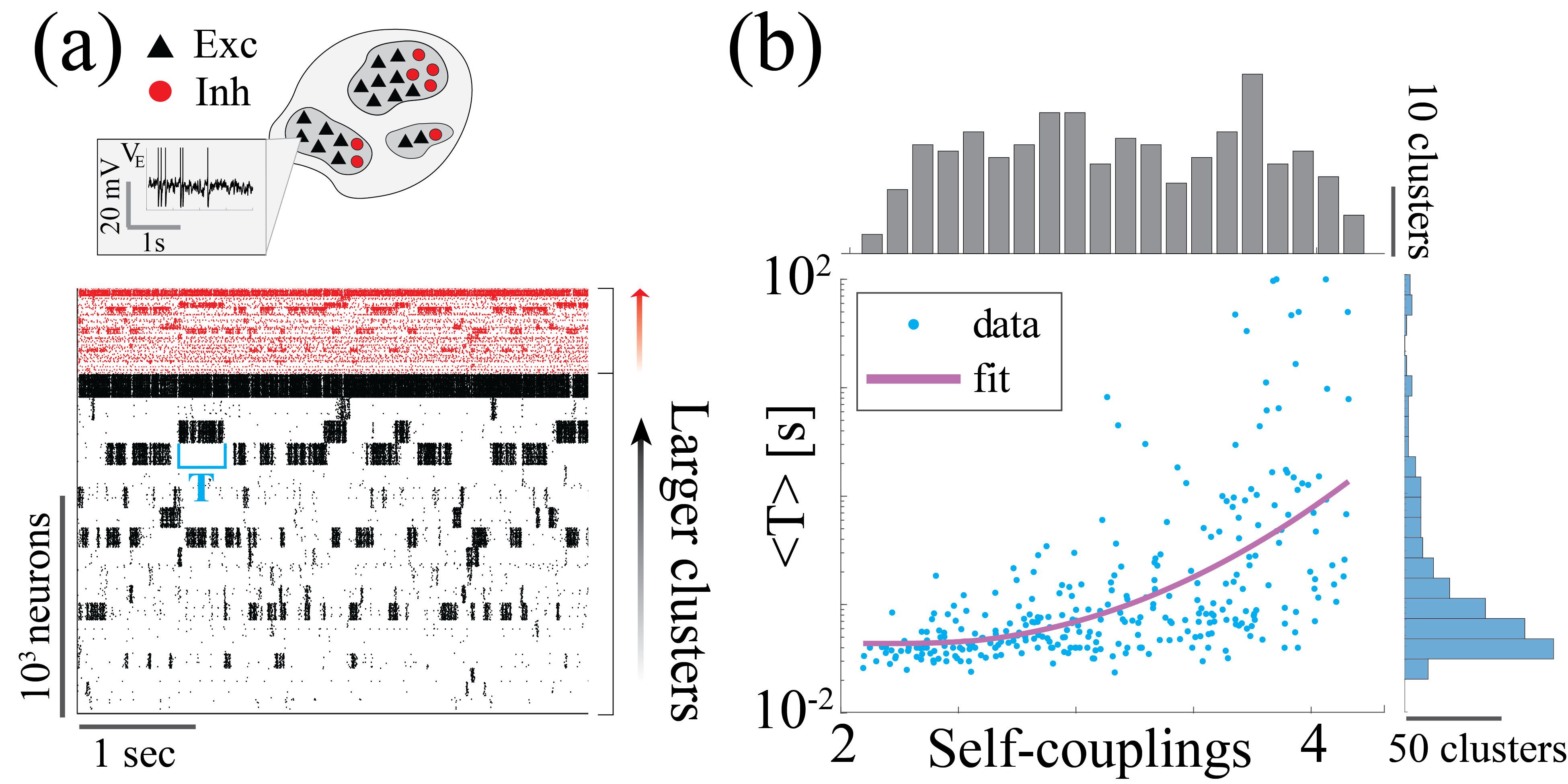}
    \caption{\lm{a) Heterogeneity of timescales in E-I spiking networks. Top: Schematic of a spiking network with excitatory (black) and inhibitory populations (red) arranged in clusters with heterogeneous distribution of sizes. Bottom: In a representative trial, neural clusters activate and deactivate at random times generating metastable activity (neurons are sorted according to cluster membership; larger clusters on top), where larger clusters tend to activate for longer intervals. b) The average activation time $T$ of a cluster increases with its self-coupling (i.e., the product of its size and average recurrent coupling), leading to a large distribution of timescales ranging from 20ms to 100s (blue dots: activation times of individual clusters from 100s simulations of 20 different networks; pink curve: fit of $\log(T)= a_2s_E^2+a_1s_E+a_0$ with $a_2=0.44,a_1=2.06,a_0=1.04$.}}
    \label{figspiking}
\end{figure}

\section{A reservoir of timescales in E-I spiking networks}

\lm{We next investigated whether the neural mechanism endowing the random neural network (\ref{dyn}) with a reservoir of timescales could be implemented in a biologically plausible model exhibiting spiking activity and excitatory/inhibitory cell-type specific connectivity. We modeled the local cortical circuit as a recurrent network of excitatory (E) and inhibitory (I) current-based leaky-integrated-and-fire neurons (see Appendix \ref{appendixspiking} for details), where both E and I populations were arranged in neural clusters (Fig. \ref{figspiking}A) \citep{Amit1997b,LitwinKumarDoiron2012,wyrick2021state}. Synaptic couplings between neurons in the same cluster were potentiated compared to those between neurons in different clusters. 
Using mean field theory, we found that the recurrent interactions of cell-type specific neurons belonging to the same cluster can be interpreted as a self-coupling, which can be expressed in terms of the underlying network parameters as $s_i^E=\bar J^{(in)}_{EE}C^E_i$, where $C^i_E$ is the cluster size and $\bar J^{(in)}_{EE}$ is the average synaptic couplings between E neurons within the cluster. The spiking network time-varying activity unfolds through sequences of metastable attractors \cite{LitwinKumarDoiron2012,wyrick2021state}, characterized by the activation of different subsets of neural clusters (Fig. \ref{figspiking}A). This metastable regime is similar to the one observed in the random neural network for large self-couplings (Fig. \ref{metastable1}) and the characteristic timescale $T$ of a cluster's metastable dynamics can be estimated from its average activation time. We tested whether the heterogeneity in the cluster self-coupling distribution could lead to a heterogeneous distribution of timescales. We endowed the clustered network with a heterogeneous distribution of cluster sizes; other sources of heterogeneity include the variability in average coupling strengths $\bar J^{(in)}$ and the Erdos-Renyi connectivity, yielding altogether a heterogeneous distribution of self-couplings (Fig. \ref{figspiking}C). We found that the cluster activation timescales $T$ varied across clusters spanning a large range from 20ms to 100s (Fig. \ref{figspiking}C). The cluster timescale $T$ was proportional to the value of a cluster's self-coupling. In particular, the functional dependence of $\log(T)$ vs. self-coupling $s_E$ was best fit by a quadratic polynomial (Fig. \ref{figspiking}C, see Appendix \ref{appendixspiking} for details), in agreement with the analytical calculation in the random network model (\ref{large_s_timescale}). We thus concluded that a reservoir of timescales can naturally emerge in biologically plausible models of cortical circuits from a heterogeneous distribution of cluster sizes. Both the range of timescales (20ms-100s) \cite{cavanagh2016autocorrelation} and the distribution of cluster sizes (50-100 neurons) \cite{Perin2011-ts,marshel2019cortical} are consistent with experimental observations.}
\begin{figure*}
\begin{centering}
\includegraphics[width=1.0\textwidth]{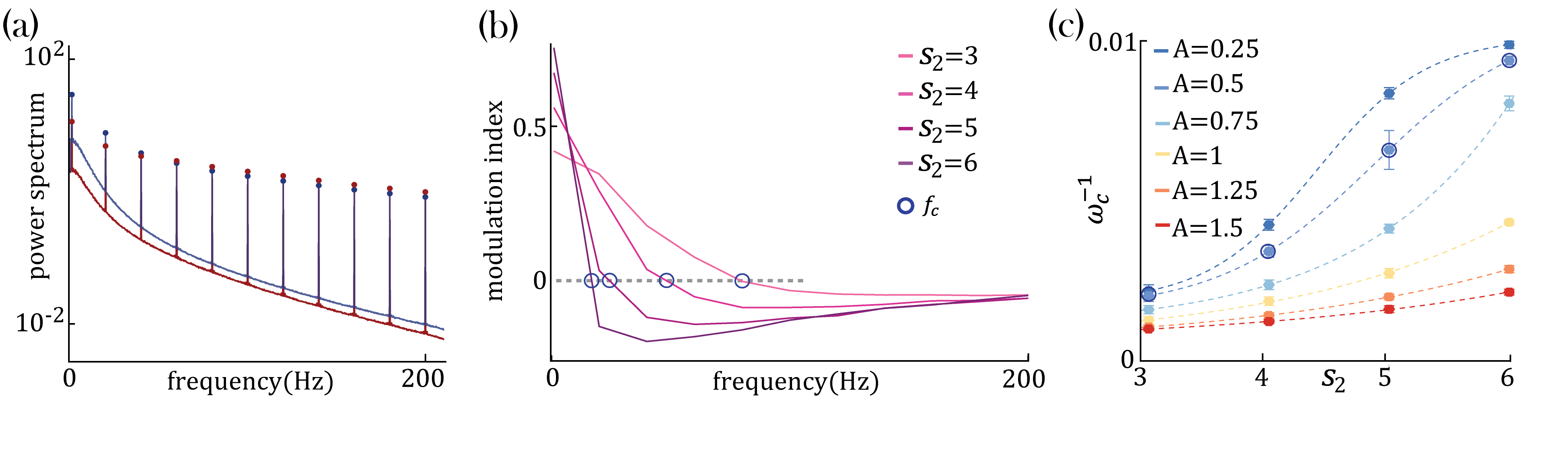}
    \caption{Network response to broadband input. (a) Power spectrum density of a network driven by time-dependent input comprising a superposition of $11$ sinusoidal frequencies (see main text for details). Maroon and navy curves represent average power spectrum density in $s_1$ and $s_2$ populations, respectively; circles indicate the peak in the power spectrum density amplitudes at each frequency; amplitude A = $0.5$; $N_{1} =N_{2} = 1000$, $g=3.0$, $s_1=1$ and $s_2=4$. (b) Modulation index, Eq.~(\ref{eq:modulation_index}), of the power spectrum density amplitudes as a function of frequency in networks with $s_1 = 1$ and various $s_2$. The green circles mark the cutoff frequency $f_{c}$ where the modulation index changes sign. (c) Cutoff period, $2\pi \omega_c^{-1} $, as a function of self coupling $s_2$ for different input amplitudes. An inversely proportional relation between the cut off period and the amplitude of the broadband signal is present.}
    \label{figthree}
\end{centering}
\end{figure*}

\section{Demixing of time-dependent broadband input}
Previous work in random networks with no self-couplings ($s_i=0$ in (\ref{dyn})) showed that stimulus-driven suppression of chaos is enhanced at a particular input frequency, related to the network's intrinsic timescale \cite{rajan2010stimulus}. We investigated whether in our network with two different self-couplings $s_1<s_2$, in the chaotic regime, the stimulus-dependent suppression of chaos exhibited different features in the two sub-populations, depending on their different intrinsic timescale. We drove each network unit $x_i$ with an external broadband stimulus $I_i(t)=A\sum_{l=1}^L\sin(2 \pi f_l t+\theta_i)$ consisting of the superposition of $L$ sinusoidal inputs of different frequencies $f_l$ in the range $1-200$ Hz, with an equal amplitude $A=0.5$ and random phases $\theta_i$. We found that the sub-population with a slow, or fast, intrinsic timescale preferentially entrained its activity with slower, or faster, spectral components of the broadband stimulus respectively (Fig. \ref{figthree}a). We quantified this effect using a spectral modulation index 
\begin{eqnarray}
m(f)=[(P_2(f)-P_1(f))/(P_2(f)+P_1(f))] \,
\label{eq:modulation_index}
\end{eqnarray}
where $P_\alpha(f)$ is the power-spectrum peak of sub-population $\alpha$ at the frequency $f$ (Fig. \ref{figthree}b). A positive, or negative, value of $m(f)$ reveals a stronger, or weaker, respectively, entrainment at frequency $f$ in the sub-population $s_2$ compared to $s_1$. $m(f)$ exhibited a crossover behavior whereby the low frequency component of the input predominantly entrained the slow population $s_2$, while the fast component of the input predominantly entrained the fast population $s_1$. When fixing $s_1=1$ and varying $s_2$, we found that the dependence of the crossover frequency $f_c$ on $s_2$ was strong at low input amplitudes and was progressively tamed at larger input amplitudes (Fig. \ref{figthree}c). This is consistent with the fact that the input amplitude completely suppresses chaos beyond a certain critical value, as previously reported in network's with no self-couplings \cite{rajan2010stimulus}. 

\section{Discussion}

\lm{We demonstrated a new robust and biologically plausible network mechanism whereby multiple timescales emerge across units with heterogeneous self-couplings. In our model, units are interpreted as neural clusters, or functional assemblies, consistent with experimental evidence from cortical circuits  \citep{Perin2011-ts,Lee2016-ld,kiani2015natural,miller2014visual,marshel2019cortical} and theoretical modeling \cite{LitwinKumarDoiron2012,wyrick2021state}. We found that the neural mechanism underlying the large range of timescales is the heterogeneity in the distribution of self-couplings (representing neural cluster strengths). We showed that this mechanism can be naturally implemented in a biologically plausible model of a neural circuit based on spiking neurons with excitatory/inhibitory cell-type specific connectivity. This spiking network represents a microscopic realization of our mechanism where neurons are arranged in clusters and a cluster's self-coupling represents the strength of the recurrent interactions between neurons belonging to that cluster. A heterogeneous distribution of cluster sizes, in turn, generates a reservoir of timescales. 
}

\lm{Several experimental studies uncovered heterogeneity of timescales of neural activity across brain areas and species. Comparison of the population-averaged autocorrelations across cortical areas revealed a hierarchical structure, varying from 50ms to 350ms along the occipital-to-frontal axis \citep{murray2014hierarchy}. Neurons within the same area exhibit a wide distribution of timescales as well. A heterogeneous distribution of timescales (from 0.5s to 50s) was found across neurons in the oculomotor system of the fish \citep{miri2011spatial} and primate brainstem \citep{joshua2013diversity}, suggesting that timescale heterogeneity is conserved across phylogeny. During periods of ongoing activity, the distribution of single-cell autocorrelation timescales in primates was found to be right-skewed and approximately lognormal, ranging from 10ms to 10s \citep{cavanagh2016autocorrelation}. Single neuron activity was found to encode long reward memory traces in primate frontal areas over a wide range of timescales up to 10 consecutive trials  \citep{bernacchia2011reservoir}. In these studies, autocorrelation timescales where estimated using parametric fits, which may be affected by statistical biases, although a new Bayesian generative approach might overcome this issue \cite{zeraati2020estimation}. In this study, we estimated timescales nonparametrically as the half-width at half-maximum of the autocorrelation function. In our biologically plausible model based on a spiking network with cell-type specific connectivity, the distribution of timescales was in the range between 20ms and 100s, similar to the range of timescales observed in experiments \citep{miri2011spatial,joshua2013diversity,cavanagh2016autocorrelation}. Moreover, the distribution of cluster sizes in our model is within the 50-100 neurons range, consistent with the size of functional assemblies experimentally observed in cortical circuits \cite{Perin2011-ts,marshel2019cortical}. A fundamental new prediction of our model, to be tested in future experiments, is the direct relationship between cluster strength and its timescale.
}

\lm{Previous neural mechanisms for generating multiple timescales of neural activity relied on single cell bio-physical properties, such as membrane or synaptic time constants  \citep{gjorgjieva2016computational}. In feedforward networks, developmental changes in single-cell conductance can modulate the timescale of information transmission, explaining the transition from slow waves to rapid fluctuations observed in the developing cortex \cite{gjorgjieva2014intrinsic}. However, the extent to which this single-cell mechanism might persist in presence of strong recurrent dynamics was not assessed. To elucidate this issue, we examined whether a heterogeneous distribution of single-unit integration time constants could lead to a separation of timescales in a random neural network (see Appendix D for details). In this model, half of the units were endowed with a fast time constant which we held fixed, and the other half with a slow time constant, whose value we varied across networks. We found that, although the average network timescale increased proportionally to the value of the slower time constants, the difference in autocorrelation time between the two populations remained negligible. These results suggest that, although the heterogeneity in single-cell time constants may affect the dynamics of single neurons in isolation or within feedforward circuits \cite{gjorgjieva2014intrinsic}, the presence of strong recurrent dynamics fundamentally alter these single-cell properties in a counterintuitive way. Our results suggest that a heterogeneity in single cell time constants may not lead to a diversity of timescales in presence of recurrent dynamics.}

\lm{Our results further clarified that the relationship between a cluster's self-coupling and its timescale relies on the strong recurrent dynamics. This relationship is absent when driving an isolated cluster with white noise external input (Fig. \ref{metastable1}). Indeed, the mechanism linking the self-coupling to the timescale only emerged when driving the unit with a mean field whose color was self-consistently obtained from an underlying recurrent network of self-coupled units.   
}

\lm{
Previous models showed that a heterogeneity of timescales may emerge from circuit dynamics through a combination of structural heterogeneities and heterogeneous long-range connections arranged along a spatial feedforward gradient \cite{chaudhuri_xjw2014diversity,chaudhuri2015large}. These networks can reproduce the population-averaged hierarchy of timescales observed across cortex in the range of 50-350ms \cite{murray2014hierarchy,chaudhuri2015large}. A similar network architecture can also reproduce the heterogeneous relaxation time after a saccade, found in the brainstem oculomotor circuit \cite{miri2011spatial,joshua2013diversity}, in a range between 10-50s \cite{inagaki2019discrete,recanatesi2022metastable}. This class of models can explain a timescale separation within a factor of 10, but it is not known whether they can be extended to several orders of magnitude, as observed between neurons in the same cortical area \cite{cavanagh2016autocorrelation}. Moreover, while the feedforward spatial structure underlying these two models is a known feature of the cortical hierarchy and of the brainstem circuit, respectively, it is not known whether such a feedforward structure is present within a local cortical circuit. Our model, on the other hand, relies on strong recurrent connectivity and local functional assemblies, two hallmarks of the architecture of local cortical circuits \cite{Perin2011-ts,Lee2016-ld,kiani2015natural,miller2014visual,marshel2019cortical}. Other network models generating multiple timescales of activity fluctuations  were  proposed based on self-tuned criticality with anti-hebbian plasticity \cite{magnasco2009self}, or multiple block-structured connectivity \cite{merav2015blocks}.
}

In our model, the dynamics of units with large self-couplings, exhibiting slow switching between bistable states, can be captured analytically using the universal colored noise approximation to the Fokker-Planck equation \cite{hanggi1995colored,jung1987dynamical}. This is a classic tool from the theory of stochastic processes, which we successfully applied to investigate neural network dynamics for the first time. This slow switching regime may underlie the emergence of metastable activity, ubiquitously observed in the population spiking activity of behaving mammals \cite{Abeles1995a,Jones2007,mazzucato2015dynamics,mazzucato2019expectation,recanatesi2022metastable,engel2016selective,kadmon2019movement}.

What is the functional relevance of neural circuits exhibiting a reservoir of multiple timescales? The presence of long timescales deeply in the chaotic regime is a new feature of our model which may be beneficial for memory capacity away from the edge of chaos \cite{toyoizumi2011beyond}. Moreover, we found that, in our model, time-dependent broadband inputs suppress chaos in a population-specific way, whereby slow (fast) subpopulations preferentially entrain slow (fast) spectral components of the input. This mechanism may thus endow recurrent networks with a natural and robust tool to spatially demix complex temporal inputs \cite{perez2021neural} as observed in visual cortex \cite{mazzoni2008encoding}. Third, the presence of multiple timescales may be beneficial for performing flexible computations involving simultaneously fast and slow timescales, such as in role-switching tasks \cite{iigaya2019deviation}; or as observed in time cells in the hippocampus \cite{kraus2013hippocampal,howard2014unified}. A promising direction for future investigation is the exploration of the computational properties of our model in the context of reservoir computing \cite{sussillo2009generating} or recurrent networks trained to perform complex cognitive tasks \cite{yang2019task}.

\acknowledgments

We would like to thank G. Mongillo and G. La Camera for discussions. LM was supported by National Institute of Neurological Disorders and Stroke grant R01-NS118461 and by National Institute on Drug Abuse grant R01-DA055439 (CRCNS). MS was supported by The Hebrew University of Jerusalem "Emergency response to covid19" grant.

\bibliography{bib} 

\begin{appendices}
\section{Dynamical regions of networks with identical self-couplings, a summary}

It is constructive to quickly survey the results of Stern et al.  \cite{merav2014clusters} who studied the special case of including a single value self-coupling $s$ for all clusters in the network, $P(s_i) = \delta_{s,s_i}$. In this case the dynamics of all units in the network follow:
\begin{equation}
\frac{dx_i}{dt} = -x_i + s \tanh(x_i) + g \sum_{i=1}^{N} J \phi(x_j),
\label{eq:dyn}
\end{equation}

Two variables determine the network dynamics, the network gain $g$ and the self-coupling value $s$. 
The network gain $g$ defines the strength of the network impact on its units. 
It brings the network into chaotic activity, without self-coupling ($s=0$), for values $g>1$ \cite{sompolinsky1988chaos}. 
The self-coupling $s$ generates bi-stability. Without network impact ($g=0$) the dynamical equation \ref{eq:dyn} for each unit becomes
\begin{equation}
\frac{dx_i}{dt} = -x_i + s \tanh(x_i),
\label{eq:dyn_single_unit}
\end{equation}
which has two stable solutions for $s>1$ (Appendix Fig. 1a), both at $x \neq 0$. For $s<1$ (Appendix Fig. 1b) a single stable solution exists at $x=0$.

When small values of network gain $g$ are introduced to the network dynamics, Eq.~(\ref{eq:dyn}), with identical bi-stable units ($s>1$), each unit solution jitters around one of its two possible fixed points. After an irregular activity the network settle into a stable fixed point. 
This generates a region of transient irregular activity with stable fixed points (Appendix Fig. 1c). 
As $g$ increases and $s$ decreases, different possible fixed point configurations lose their stability (as a result, the typical time spent in the transient activity increases). 
When the critical line $s_c \approx 1 + 0.157 \ln{(0.443g + 1)}$ is crossed, no fixed point remains stable and the network activity becomes chaotic \cite{merav2014clusters}. 
The ``last" stable fixed point at the transition line has a unique configuration with all unit values located farthest from $x=0$ (Appendix Fig. 1a, light green lines). Additional decrease of $s$ and $g$ leads to a region where any initial activity of the network decays and the trivial solution ($x_i=0$ for all $i$) is stable (Appendix Fig. 1c).

\begin{figure}
  \centering
  \label{fig:identical_selfcoupling}
    \includegraphics[width=1.0\linewidth]{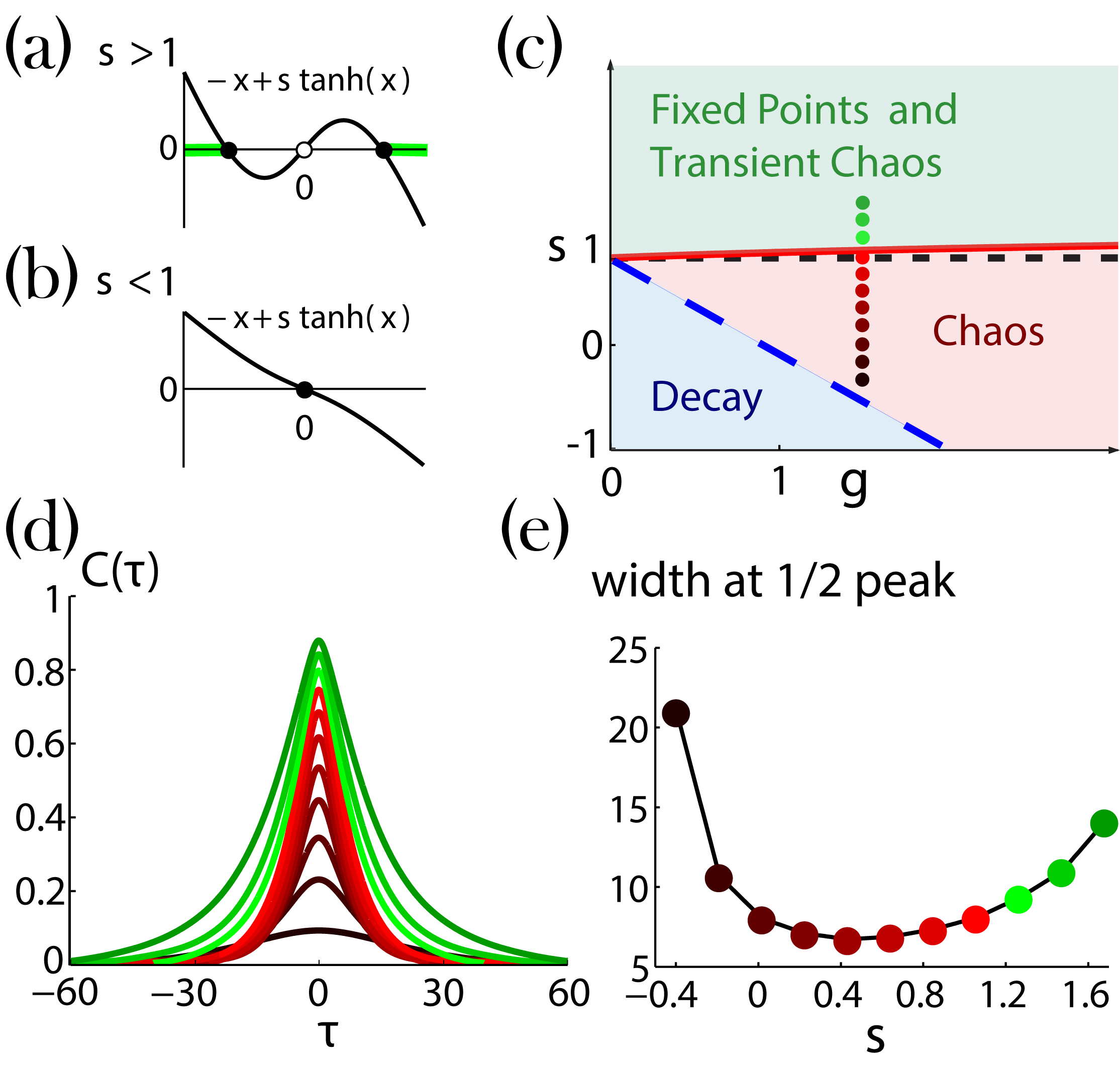}
  \caption{Network dynamics with identical self-couplings, adopted from \cite{merav2014clusters}.  
  a,b) Graphical solutions to Eq.~(\ref{eq:dyn_single_unit}). a) For $s>1$ there are two stable non zero solutions (full black circles) and an unstable solution at zero (open black circle). The green background over the $x$ axis denotes the regions of allowed activity values at the stable fixed point on the transition line to chaos (solid red curve in (c)). b) For $s<1$ there is a single stable solution (full black circle) at zero. c) Regions of the network dynamics over a range of $s$ and $g$ values. Below the long dashed blue line any initial activity in the network decays to zero. Above the solid red curve, the network exhibits transient irregular activity that eventually settles into one out of a number of possible nonzero stable fixed points. In the region between these two curves, the network activity is chaotic. Colored circles denote, according to their locations on the phase diagram and with respect to their colors, the values of $s$ (ranging from $1.6$ and decreasing with steps of $0.2$) and $g = 1.5$, used for the autocorrelation functions $C(\tau)$ in (d)~\footnote{Corrected version}. e) Widths at half peak (values of $\tau$'s in the main text notation) of the autocorrelation functions in (d).}
\end{figure}

\end{appendices}
\begin{appendices}
\section{Mean field theory with multiple self-couplings}

We derive the dynamic mean-field theory in the limit $N\rightarrow\infty$ by using the moment generating functional~\cite{Sompolinsky1982EAmodel,Sompolinsky1987langevin}. For the derivation we follow the Martin-Siggia-Rose-De Dominicis-Janssen path integral approach formalism ~\cite{MSR1973dmft} as appears extensively in~\cite{helias2020statistical}, we borrow their notations as well. For our model, Eq.~(\ref{dyn}), the moment generating functional is given by:
\begin{equation}
 \label{moment_generating_functional}
 \begin{split}
Z =  \int \mathcal{D}\tilde{x} \mathcal{D} x \exp \Bigg[&\int dt \sum_{i=1}^N\tilde{x}_i(t) \big[ (\partial_{t} + 1 ) x_{i}(t) -s_{i}\phi(x_i(t))\big] \\
&+\sum_{i=1}^N\lambda_i(t)x_i(t)- \sum_{j\neq i}\tilde{x}_i(t) J_{ij}\phi(x_j(t))  \Bigg] ,
\end{split}
\end{equation}
where $\mathcal{D} x = \prod_{i}\mathcal{D} x_i$ and $\mathcal{D}\tilde{x} = \prod_{i}\mathcal{D}  \tilde{x}_i/2\pi i$. 
To start, we calculate $\langle Z(J) \rangle_J$. We take advantage of the self-averaging nature of our model, particularly by averaging over the quenched disorder, $J$. The couplings, ${{J_{ij}}}$, are i.i.d. variables extracted from a normal distribution and appear only in the last term in (\ref{moment_generating_functional}). We hence focus our current calculation step on that term, and we derive the result to the leading term in $N$, yielding:
\begin{equation}\footnotesize
\begin{split}
    &\int \prod_{i\neq j}dJ_{ij}\sqrt{\frac{N}{2\pi g^2}}  \exp{\left[ - \frac{J_{ij}^2 N}{2g^2}  \right]}\exp{\big[  -\int dt\,\tilde x_i(t) J_{ij}\phi(x_j(t)) \big] } \\
            &= \exp{\left[\frac{1}{2} \int dtdt' \left(\sum_i \tilde{x}_{i}(t)\tilde{x}_{i}(t')\right) \left( \frac{g^2}{N}\sum_j\phi(x_j(t))\phi(x_j(t')) \right)\right]}.
\end{split}
\label{MGF_avg}
\end{equation}   
The result above suggests that all the units in our network are coupled to one another equivalently (by being coupled only to sums that depend on all units' activity). To further decouple the network, we define the quantity 
$$
Q_1(t,t') \equiv \frac{g^2}{N} \sum_j \phi(x_j(t))\phi(x_j(t')).
$$
We enforce this definition by multiplying the disordered averaged moment generating functional with the appropriate Dirac delta function, $\delta$, in its integral form:
\begin{equation*}
\begin{split}
1 = & \int dQ_1 \delta \big[ -\frac{N}{g^2} Q_1 + \sum_j \phi(x_j(t))\phi(x_j(t')) \big] \\
 = & \int dQ_1 dQ_2 \exp{Q_2 \big[-\frac{N}{g^2} Q_1 + \sum_j \phi(x_j(t))\phi(x_j(t'))\big]},
\end{split}
\end{equation*}
where $dQ_2$ is an integral over the imaginary axis (including its $1/ (2\pi i)$ factor). 
We can now rewrite the disordered averaged moment generating functional, using (\ref{MGF_avg}) to replace its last term, the definition of $Q_1$, and with multiplying the functional by the $\delta$ function above. All together we get:
\begin{equation}\small
\begin{split}
    \langle Z(J) \rangle_J &= \int dQ_1dQ_2 \exp\Big[ -\frac{N}{g^2} \int dt dt' Q_1Q_2 \Big. \\
                           &+ \Big. N\sum_{\alpha\in A}n_{\alpha} \ln[Z_{\alpha}] \Big] ,    \\
                Z_{\alpha} &= \int \mathcal{D} \tilde{x}_{\alpha}\mathcal{D} x_{\alpha} \exp\Bigg[ \int dt \tilde{x}_{\alpha}(t) \Big((\partial_t + 1)x_{\alpha}(t) \Big. \Big.\\
    &- \Big. s_{\alpha}\phi(x_{\alpha}(t))\Big)\\
    & + \frac{1}{2}\int dt dt' \tilde{x}_{\alpha}(t)Q_1(t,t')\tilde{x}_{\alpha}(t')  \\
    &+ \Big. \int dt dt' {\phi}(x_{\alpha}(t))Q_2(t,t'){\phi}(x_{\alpha}(t'))\Bigg],
\end{split}
\label{short_MGF_decoupled}
\end{equation}
with $n_\alpha=N_\alpha/N$ the fraction of units with self-couplings $s_\alpha$ across the population, for $\alpha \in A$. 
In the expression above we made use of the fact that $Q_1$ and $Q_2$, now in a role of auxiliary fields, couple to sums of the fields $x_i^2$ and $\phi_i^2$ and hence the generating functional for $x_i$ and $\tilde x_i$ can be factorized with identical multiplications of $Z_\alpha$. Note that in our network, due to the dependency on $s_i$, $x_i$-s are equivalent as long as $s_i$-s are equivalent. Hence, the factorization is for $Z_\alpha$ for all $x_i$ with $s_i=s_\alpha$. Now each factor $Z_\alpha$ includes the functional integrals $\mathcal{D} x_\alpha$ and $\mathcal{D} \tilde x_\alpha$ for a single unit with self-coupling $s_\alpha$. 

In the large $N$ limit we evaluate the auxiliary fields in  (\ref{short_MGF_decoupled}) by the saddle point approach (we note variable valued at the saddle point by $*$),  obtaining:
\begin{equation*}
0 = \frac{\delta}{\delta Q_{1,2}}\left[ -\frac{1}{g^2} \int dt dt' Q_1Q_2 + \sum_{\alpha \in A }n_{\alpha} \ln[Z_{\alpha}] \right] \ ,
\end{equation*}
and yielding the saddle point values $(Q_1^*,Q_2^*)$:
\begin{equation}
\begin{split}
    0 &= - \frac{1}{g^2} Q_1^* (t,t') +\sum_{\alpha \in A}\frac{n_{\alpha}}{Z_{\alpha}} \frac{\partial Z_{\alpha}}{\partial Q_2(t,t')}\Bigg\vert_{Q^*} \\
    &~\Leftrightarrow Q_1^*(t,t') = g^2\sum_{\alpha \in A }n_\alpha\langle\phi(x_{\alpha}(t))\phi(x_{\alpha}(t'))\rangle \equiv g^2 C(\tau),
\end{split}
\label{C_phi}
\end{equation}

\begin{equation}
\begin{split}
    0 &= -\frac{1}{g^2}Q_2^* (t,t') +\sum_{\alpha \in A }\frac{n_{\alpha}}{Z_{\alpha}}\frac{\partial Z_{\alpha}}{\partial Q_1(t,t')}\Bigg\vert_{Q^*} \\
    &~\Leftrightarrow Q_2^*(t,t') = \frac{g^2}{2}\sum_{\alpha \in A }n_\alpha\langle\tilde{x}_{\alpha}(t)\tilde{x}_{\alpha}(t')\rangle = 0,
\end{split}
\end{equation}
where $C(\tau)$, with $\tau=f(t,t')$, represents the average autocorrelation function of the network (as was defined in the main text). The second saddle point $Q_2^* = 0$ vanishes due to $\langle\tilde{x}_\alpha(t)\tilde{x}_\alpha(t')\rangle=0$ as can be immediately extended from  \cite{helias2020statistical,Sompolinsky1982EAmodel}.
The action at the saddle point reduces to the sum of actions for individual, non-interacting units with self-coupling $s_\alpha$. All units are coupled to a common external field $Q_1^*$. Inserting the saddle point values back into Eq.~(\ref{short_MGF_decoupled}), we obtain $Z^*=\prod_\alpha (Z^*_\alpha)^{N_\alpha}$ where
\begin{equation}
\begin{split}
    {Z}^*_\alpha \sim &\int \mathcal{D}\tilde{x}_{\alpha}\mathcal{D}x_{\alpha} \exp \sum_{\alpha \in A} \Big(  \int dt \tilde{x}_{\alpha}(t)\big((\partial_t + 1 )x_{\alpha}(t) \\
    &-s_{\alpha}\phi(x_{\alpha}(t)) \big) + \frac{g^2}{2} \int dt dt' \tilde{x}_{\alpha}(t)C(\tau)\tilde{x}_{\alpha}(t')  \Big).
\end{split}
\end{equation}
Thus in the large $N$ limit the network dynamics are reduced to those of a number of $A$ units $x_\alpha(t)$, each represents the sub-population with self-couplings $s_\alpha$ and follows dynamics governed by
\begin{equation}
   \frac{d}{dt}x_{\alpha}(t) = -x_{\alpha}(t) + s_{\alpha} \phi[x_\alpha(t)] + \eta(t) \ 
   \label{eqdmftfinal}
\end{equation}
for all $\alpha \in A$ and where $\eta(t)$ is a Gaussian mean field with autocorrelation
\begin{equation}
   \langle\eta(t)\eta(t')\rangle = g^2\sum_{\alpha \in A }n_\alpha\langle\phi(x_{\alpha}(t))\phi(x_{\alpha}(t'))\rangle.
   \label{eqdmftfinal2}
\end{equation}

The results above can be immediately extended for the continuous case of self-coupling distribution $P(s)$ yielding:
\begin{equation}
   \langle\eta(t)\eta(t')\rangle = g^2 \int p(s) \phi(x(s,t))\phi(x(s,t')) ds \,
   \label{eqdmftfinal2_continuous}
\end{equation}
with $p(s)$ the density function of the self-couplings distribution in the network and the units dynamics dependent on their respective self-couplings:
\begin{equation}
   \frac{d}{dt}x(s,t) = -x(s,t) + s \phi[x(s,t)] + \eta(t) \ .
   \label{eqdmftfinal_continuous}
\end{equation}

\section{Spiking network model}
\label{appendixspiking}

\lm{{\it Clustered network architecture. }We simulated a recurrent network of $N=2000$ excitatory (E) and inhibitory (I) spiking neurons with relative fractions $n_E=80\%$ and $n_I=20\%$ and connection probabilities $p_{EE}=0.2$ and $p_{EI}=p_{IE}=p_{II}=0.5$ (Fig.~\ref{figspiking}). Non-zero synaptic weights from pre-synaptic neuron $j$ to post-synaptic neuron $i$ were $J_{ij}={j_{ij}/\sqrt{N}}$, with $j_{ij}$ sampled from a normal distribution with mean $j_{\alpha\beta}$, for $\alpha,\beta=E,I$, and standard deviation $\delta^2$. Neurons were arranged in $p$ cell-type specific clusters. E clusters had heterogeneous sizes drawn from a uniform distribution with mean of $N^{clust}_E=80$ E-neurons and $30\%$ standard deviation. The number of clusters was determined as $p=\textrm{round}(n_EN(1-n_{bgr})/N^{clust}_E)$, where $n_{bgr}=0.1$ is the fraction of background neurons in each population, i.e., not belonging to any cluster. I clusters were paired to E clusters and the size of each I cluster was matched to the corresponding E cluster with a proportionality factor $n_I/n_E=1/4$. Neurons belonging to the same cluster had potentiated intra-cluster weights by a factor $J^{+}_{\alpha\beta}$, while those belonging to different clusters had depressed inter-cluster weights by a factor $J^{-}_{\alpha\beta}$, where: $J^{+}_{EI}=p/(1+(p-1)/g_{EI})$, $J^{+}_{IE}=p/(1+(p-1)/g_{IE})$, $J^-_{EI}=J^{+}_{EI}/g_{EI}$, $J^-_{IE}=J^{+}_{IE}/g_{IE}$ and
$J^-_{\alpha\alpha}=1-\gamma (J^+_{\alpha\alpha}-1)$ for $\alpha=E,I$, with $\gamma=f(2-f(p+1))^{-1}$. $f=(1-n_{bgr})/p$ is the fraction of E neurons in each cluster. Parameter values are in Table \ref{table:1}.}
 
\lm{{Single neuron dynamics.} We simulated current-based leaky-integrate-and-fire (LIF) neurons, with membrane potential $V$ and dynamical equation
$$
\frac{dV}{dt}=-\frac{V}{\tau_m} +I_{rec} +I_{ext} \ ,
$$
where $\tau_m$ is the membrane time constant. Input currents included a contribution $I_{rec}$ from the other recurrently connected neurons and a constant external
current $I_{ext}=N_{ext}J_{\alpha0}r_{ext}$ (units of mV s${}^{-1}$), for $\alpha=E,I$, representing afferent inputs from other brain areas and $N_{ext}=n_ENp_{EE}$. When the membrane potential $V$ hits the threshold $V^{thr}_{\alpha}$ (for $\alpha=E,I$), a spike
is emitted and $V$ is held at the reset value $V^{reset}$ for a refractory period $\tau_{refr}$. We chose the thresholds so that the homogeneous network (i.e.,where all $J^{\pm}_{\alpha\beta}=1$) was in a balanced state with average spiking activity at rates
$(r_E,r_I) = (2,5)$ spks/s. The post-synaptic currents evolved according to
$$
\tau_{syn}\frac{dI_{rec}}{dt}=-I_{rec}+\sum_{j=1}^NJ_{ij}\sum_{k}\delta(t-t_k) \ ,
$$
where $\tau_s$ is the synaptic time constant, $J_{ij}$ are the recurrent couplings and $t_k$ is the time of the k-th spike from the
j-th presynaptic neuron. Parameter values are in Table \ref{table:1}. 
}

\lm{
{\it Self-couplings.} We can estimate the E-cluster self-couplings in this model using mean field methods  \citep{Amit1997b,wyrick2021state}. The infinitesimal mean $\mu$ of the postsynaptic input to a neuron in a representative E cluster in focus is
\begin{equation}
\label{eq:mean}
\begin{split}
    &\mu_1= N n_E p_{EE}J_{EE}\Bigl[
    J^+_{EE}f^E_1r^E_1\\
    &+J_{EE}^-\sum_{l=2}^{p}f^E_l r^E_l+ n_{bg}r^{E}_{bg}\Bigr]+N_{ext}{J_{E0}}r_{ext}\\
    &- N n_{I}p_{EI}J_{EI}\Bigl[f_1^I J^+_{EI}r^I_1+J_{EI}^-\sum_{l=2}^{p}f^I_l r^I_l+ n_{bg}r^{I}_{bg} \Bigr]\ , 
    \end{split}
\end{equation}
where $r_1^E$ is the firing rate of the E cluster in focus, $r_1^I$ is the firing rate of its paired I cluster; $r^E_l,r^I_l$, for $l=2,\ldots,p$ are the firing rates of the other E and I clusters; $r^{E}_{bg},r^{I}_{bg}$ are the firing rates of the background (unclustered) populations. $f^E_i,f^I_i$ represent the fraction of E and I neurons in each cluster, which are drawn from a uniform distribution (see above). The first line in (\ref{eq:mean}) represent the contribution to the input current coming from neurons within the same E cluster, or, in other words, the self-coupling of the cluster in focus. We can thus recast the first term in the input current as $s_1r^E_1$ where $s_1=N n_E p_{EE}J_{EE}J^+_{EE}f^E_1$. The number of neurons in the cluster is given by $N_1=Nn_Ef_1^E$, and the average E-to-E synaptic coupling is $\bar J^{(in)}=p_{EE}J_{EE}J^+_{EE}$, from which we obtain $s_1=N_1\bar J^{(in)}_{EE}$, which is the expression we used in Fig. (\ref{figspiking}).  We can thus recast (\ref{eq:mean}) as 
\begin{equation}
\label{eq:mean2}
\begin{split}
    \mu_1=& s^E_1r_1^E-s^I_1r_1^I+\sum_{l=2}^p (\hat J_{1l}^{EE} r^E_l-\hat J_{1l}^{EI} r^I_l) \\
    &+\hat J^{bg,E}r^E_{bg}-\hat J^{bg,I}r^I_{bg}+\hat J^{ext}r_{ext}\ ,
    \end{split}
\end{equation}
where $\hat J$ represent effective synaptic couplings which depend on the underlying spiking network parameters in (\ref{eq:mean}). The first line in (\ref{eq:mean2}) exhibits the same functional form as the rate model in Eq. (\ref{dyn}), if we identify each rate unit as a functional cluster with a corresponding self-coupling. A crucial simplification occurring in the rate model (\ref{dyn}) is the absence of cell-type specific connectivity and the corresponding difference in the statistics of the distribution of the effective couplings $\hat J$, whose mean is zero in  (\ref{dyn}) but non-zero in (\ref{eq:mean2}).
}

\begin{table*}[!t]
\begin{tabular}{ |p{2cm}|p{10cm}||p{2cm}|  }
\hline
\multicolumn{3}{|c|}{Model parameters for clustered network simulations} \\
\hline
Parameter& Description & Value \\
\hline
$j_{EE}$ & mean E-to-E synaptic weights $\times\sqrt{N}$ & 0.6 mV\\
$j_{IE}$ & mean E-to-I synaptic weights $\times\sqrt{N}$&0.6 mV\\
$j_{EI}$ &  mean I-to-E synaptic weights $\times\sqrt{N}$&1.9 mV\\ 
$j_{II}$ & mean I-to-I synaptic weights $\times\sqrt{N}$ &3.8 mV\\
$j_{E0}$ & mean E-to-E synaptic weights $\times\sqrt{N}$ &2.6 mV\\
$j_{I0}$ & mean I-to-I synaptic weights $\times\sqrt{N}$ &2.3 mV\\
$\delta$ & standard deviation of the synaptic weight distribution&$20\%$\\
$J^{+}_{EE}$ & Potentiated intra-cluster E-to-E weight factor &14\\
$J^{+}_{II}$ & Potentiated intra-cluster I-to-I weight factor &5\\
$g_{EI}$ & Potentiation parameter for intra-cluster I-to-E weights &10\\
$g_{IE}$ & Potentiation parameter for intra-cluster E-to-I weights &8\\
$r_{ext}$ & Average baseline afferent rate to E and I neurons &5 spks/s\\
$V^{thr}_E$ & E-neuron threshold potential &1.43 mV\\
$V^{thr}_I$ & I-neuron threshold potential &0.74 mV\\
$V^{reset}$ & E- and I-neuron reset potential &0 mV\\
$\tau_m$ & E- and I-neuron membrane time constant  &20 ms\\
$\tau_{refr}$ & E- and I-neuron absolute refractory period  &5 ms\\
$\tau_s$ & E- and I-neuron synaptic time constant  &5 ms\\
\hline
\end{tabular}
\caption{Parameters for the clustered network used in the simulations.}
\label{table:1}
\end{table*}


\section{RNN with Heterogeneous Time Constants}

\nicu{
Our recurrent neural network model in Eq.~\ref{dyn}, assumes that all units share the same time constant, $\theta = 1$~ms, which measures the rate of change of a neuron's membrane potential.
We examined whether a network of units with heterogeneous time constants could give rise to multiple timescales of dynamics.
We simulated the model from Eq. (\ref{dyn}) with no self-coupling term, $s_i = 0$, with neuron-specific time constant $\theta_i$:
}

\begin{equation}
    \theta_i \frac{d}{dt}x_{i}(t) = - x_{i}(t) + g\sum_j J_{ij} \phi\left[ x_{j}(t) \right].
    \label{eq:RNNwithTimeConstats}
\end{equation}

\begin{figure*}[ht!]
\begin{center}
    \includegraphics[width=1.0\textwidth]{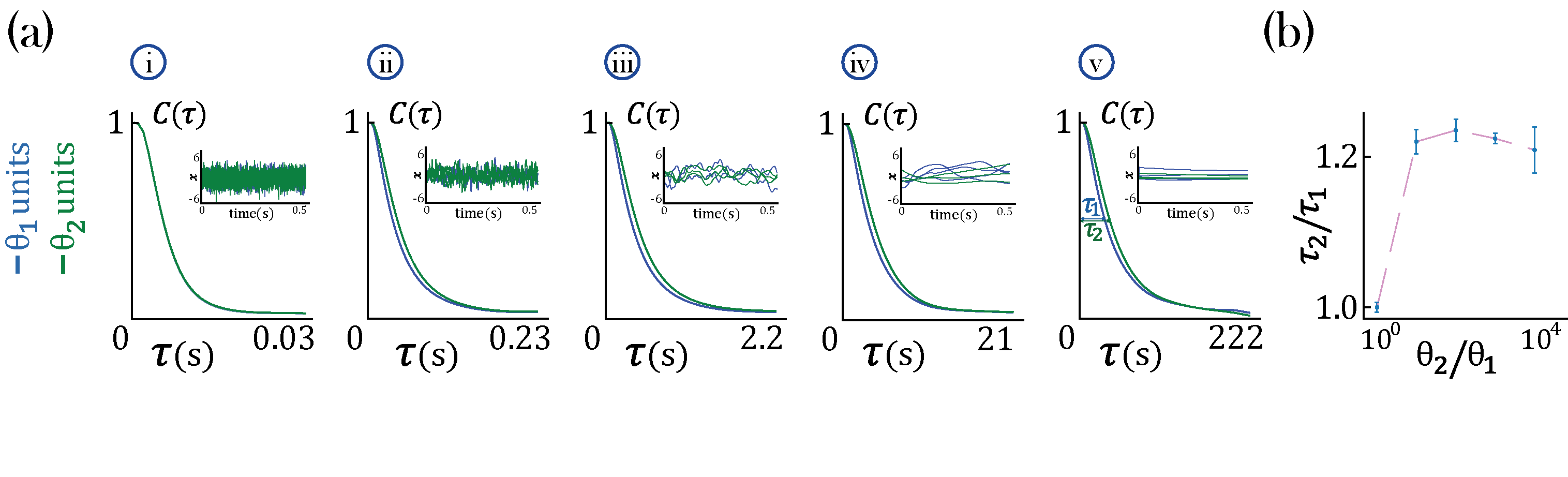}
    \caption{
    \nicu{
    Timescale analysis for an RNN with two time constants $\theta_{i}$, Eq.~\ref{eq:RNNwithTimeConstats}, governing equal populations of neurons ($N_1 = N_2 = 1000$) and gain $g = 2.5$. a) Average autocorrelation function for each population. The insert shows the dynamics of individual neurons from each population: blue for neurons with timeconstant $\theta_1$ and green for neurons with timeconstant $\theta_2$. In the networks considered here, $\theta_1 = 0.1$~ms is kept constant while: $\theta_2 = 0.1$~ms (i), $\theta_2 = 1.0$~ms (ii), $\theta_2 = 10.0$~ms (iii), $\theta_2 = 100.0$~ms (iv), $\theta_2 = 1000.0$~ms (v). b) Population timescale ratio  $\tau_2/ \tau_1$ for fixed timeconstant $\theta_1 = 0.1$~ms and varying $\theta_2$.
    }
    }
    \label{fig:RNN_with_TimeConstants}
\end{center}
\end{figure*}

\nicu{
Following the same strategy as in Fig. \ref{figone_p2}, we consider the scenario when our network contains two equal-size populations of neurons ($N_1 = N_2$) with different time constants $\theta_1\neq\theta_2$. We quantified each unit's timescales, $\tau_i$, as the width of the autocorrelation function at midpoint. When keeping $\theta_1$ fixed and increasing $\theta_2$, we found that both populations increased their timescale Fig.~\ref{fig:RNN_with_TimeConstants}a(i-v), and the ratio between the timescales of the two populations, $\tau_2 / \tau_1$ did not appreciably change over a large range time constant ratios $\theta_2/\theta_1$, Fig.~\ref{fig:RNN_with_TimeConstants}b.
}

\nicu{
Hence, we conclude that heterogeneity in single-cell time constants do not lead to large separation of timescales in network with strong recurrent dynamics.
}

\end{appendices}

\end{document}